\documentclass[prb,reprint,a4paper,superscriptaddress,showpacs,citeautoscript,floatfix]{revtex4-1}

\pdfoutput=1
\usepackage[charter]{mathdesign}
\usepackage{amsmath}

\usepackage[pdftex]{graphicx}
\usepackage{microtype}
\usepackage{bm}\let\vec\bm

\usepackage[svgnames]{xcolor}
\usepackage[
	colorlinks=True,linkcolor=DarkRed,citecolor=ForestGreen,urlcolor=MediumBlue,
	pdfstartview=FitH,bookmarks=False,pdfpagemode=UseNone
]{hyperref}

\begin{document}

\title{\boldmath Revisiting the vortex-core tunnelling spectroscopy in YBa$_2$Cu$_3$O$_{7-\delta}$}

\author{Jens Bru{\'e}r}
\author{Ivan Maggio-Aprile}
\author{Nathan Jenkins}
\author{Zoran Risti{\'c}}
\affiliation{Department of Quantum Matter Physics (DQMP), University of Geneva, 24 quai Ernest-Ansermet, 1211 Geneva 4, Switzerland}
\author{Andreas Erb}
\affiliation{Walther-Meissner-Institut, Bayerische Akademie der Wissenschaften, Walther-Meissner-Strasse 8, D-85748 Garching, Germany}
\author{Christophe Berthod}
\author{{\O}ystein Fischer}
\author{Christoph Renner}
\email[To whom correspondence should be addressed. E-mail: ]{christoph.renner@unige.ch}
\affiliation{Department of Quantum Matter Physics (DQMP), University of Geneva, 24 quai Ernest-Ansermet, 1211 Geneva 4, Switzerland}

\begin{abstract}

The observation by scanning tunnelling spectroscopy (STS) of Abrikosov vortex cores in the high-temperature superconductor YBa$_2$Cu$_3$O$_{7-\delta}$ (Y123) has revealed a robust pair of electron-hole symmetric states at finite subgap energy. Their interpretation remains an open question because theory predicts a different signature in the vortex cores, characterised by a strong zero-bias conductance peak. We present STS data on very homogeneous Y123 at 0.4~K revealing that the subgap features do not belong to vortices: they are actually observed everywhere along the surface with high spatial and energy reproducibility, even in the absence of magnetic field. Detailed analysis and modelling show that these states remain unpaired in the superconducting phase and belong to an incoherent channel which contributes to the tunnelling signal in parallel with the superconducting density of states.

\end{abstract}

\maketitle

\section{Introduction}

\begin{figure}[b]
\includegraphics[width=\columnwidth]{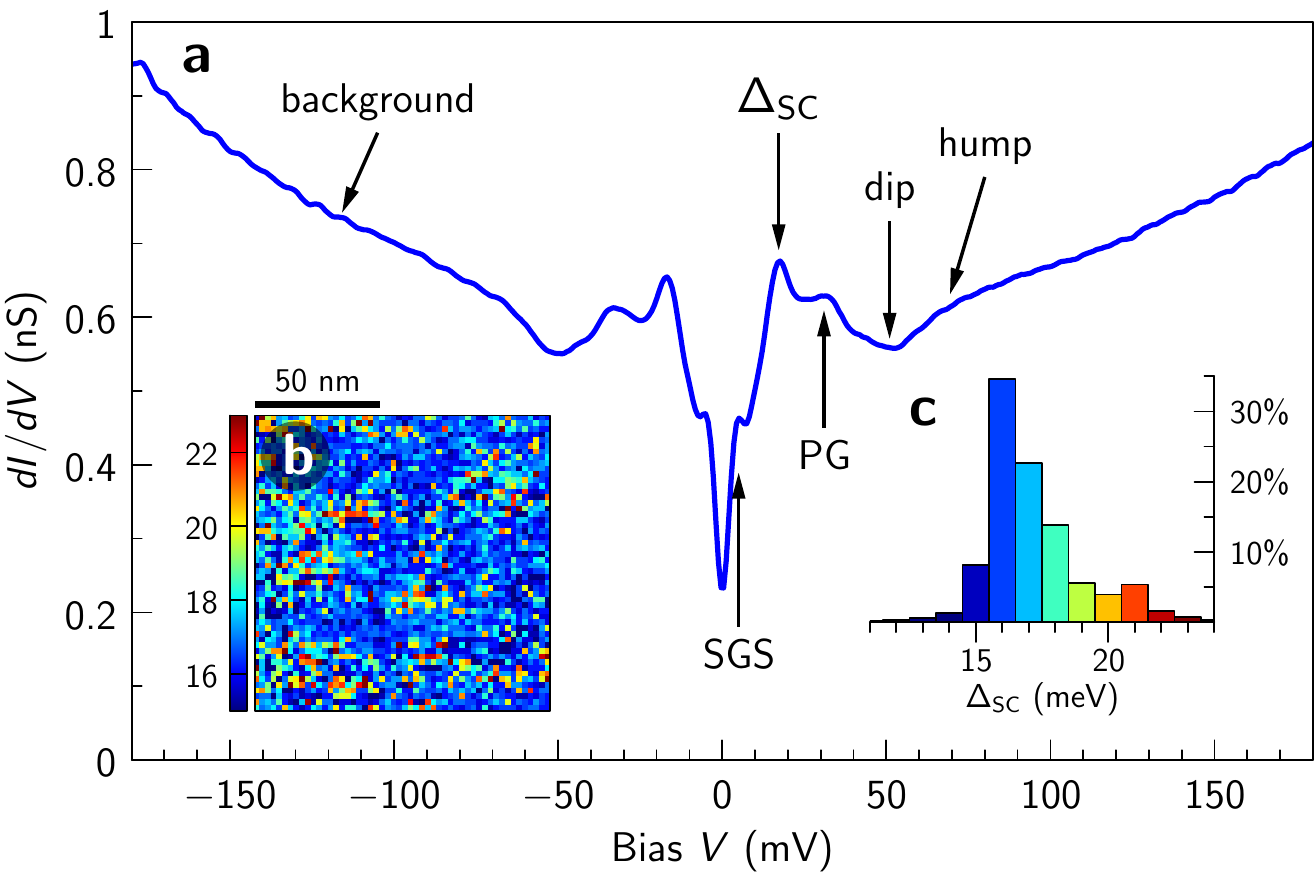}
\caption{\label{fig1} \textbf{Subgap states are observed in zero field.} (\textbf{a}) Average $dI/dV$ conductance of a $120\times120$~nm$^2$ area at $T=0.4$~K in zero field. The junction resistance was 1.2~G$\Omega$ and the regulation voltage 300~mV. Five electron-hole symmetric spectral features and the background are indicated. (\textbf{b}) Spatial map of the superconducting gap $\Delta_{\mathrm{SC}}$ over the same area. (\textbf{c}) Histogram of the gap values shown in \textbf{b}. The colouring of bars corresponds to the color scale shown in \textbf{b}.}
\end{figure}

The interaction responsible for high-temperature superconductivity in the cuprates has remained elusive until now. Various charge orders \cite{Hoffman-2002, Lawler-2010, Ghiringhelli-2012, Chang-2012, Comin-2014} suggest that the electron matter in these compounds experiences competing interactions, while numerous spectroscopic data support the scenario of preformed pairs gaining coherence at low temperature \cite{Renner-1998a, Kanigel-2008, Yang-2008, Mishra-2014, Kondo-2015}. The vortices offer a chance to disentangle the spectral features related to pairing from those that are unrelated \cite{Fischer-2007}. In this context the observation of discrete states in the vortex cores \cite{Maggio-Aprile-1995, Renner-1998b, Hoogenboom-2000a, Shibata-2010, Levy-2005, Yoshizawa-2013} has been a considerable challenge for theory \cite{Arovas-1997, Franz-1998b, Andersen-2000, Kishine-2001, Berthod-2001b, Zhu-2001a, Tsuchiura-2003, Fogelstrom-2011}, because the calculated tunnelling spectrum in a vortex of $d_{x^2-y^2}$ symmetry presents a broad continuous maximum centred at zero bias not observed in the experiments.

Here we present high-resolution scanning tunnelling microscopy (STM) measurements of YBa$_2$Cu$_3$O$_{7-\delta}$ (Y123), the first high-temperature superconductor to reveal a pair of electron-hole symmetric subgap conductance peaks when tunnelling into a vortex core \cite{Maggio-Aprile-1995}. This cuprate was chosen to revisit the vortex core spectroscopy by STM because the subgap peaks are significantly stronger than analogous low-energy spectroscopic structures reported for other cuprates \cite{Levy-2005, Yoshizawa-2013}. It furthermore shows spatially homogeneous tunnelling spectroscopy, paramount to consistently map the spatial extent of the subgap states in the vicinity of vortex cores. Our study shows that these discrete states are not a specific signature of the vortex cores, but are ubiquitous in zero field, even in spectra lacking superconducting coherence peaks. These results reshuffle the cards for the vortex core description in Y123 and shed new light on its low-temperature non-superconducting state.

\section{Results}

\subsection{Persistence of subgap states in zero field and outside vortices}

We study the low-energy spectral features of Y123 in very homogeneous conductance maps measured along the (001) surface at 0.4~K. Figure~\ref{fig1} shows the average of 2704 spectra recorded over a $120\times120$~nm$^2$ area in zero field. The persistence of sharp features in this average demonstrates the low variability of the spectral properties across the area, as well as the low noise level of the instrument. On top of a V-shaped background with a finite zero-bias conductance, which are common on as-grown Y123 surfaces, we identify five robust particle-hole symmetric spectroscopic features as labeled in Fig.~\ref{fig1}. We shall not discuss here the hump, the dip, and the pseudogap (PG), but focus on the superconducting coherence peaks (SC) near $\pm17$~meV, and especially on the subgap states (SGS) at $\pm5$~meV. We confirm the surface homogeneity by studying the spatial and statistical distributions of the superconducting gap $\Delta_{\mathrm{SC}}$, defined as half the energy separation between the two SC peaks (Fig.~\ref{fig1}b and c). The average gap is 17.2~meV with a standard deviation of 1.9~meV, corresponding to a ratio $2\Delta_{\mathrm{SC}}/k_{\mathrm{B}}T_{\mathrm{c}}=4.34\pm0.5$ where $k_{\mathrm{B}}$ is the Boltzmann constant and $T_{\mathrm{c}}$ is the critical temperature. This gap uniformity in Y123 contrasts with the case of Bi-based cuprates, where up to 50\% inhomogeneity is often observed in the gap values, even at optimal doping and for samples with sharp superconducting transitions. Most remarkable in Fig.~\ref{fig1} are the subgap states at $\pm5$~meV. Until now this structure had only been clearly resolved as peaks inside the vortices \cite{Maggio-Aprile-1995, Shibata-2010}. Much weaker subgap shoulders were occasionally observed in zero field or outside vortices. The natural interpretation has been that the structure would be strongly enhanced in the vortex cores, suggestive of a phenomenon competing with superconductivity. Our measurements show that this is not the case: the structure is as strong in zero field as it is in the vortex cores. The well-defined subgap peaks in the average spectrum show that the energy of this structure does not vary appreciably along the surface.

\begin{figure}[b]
\includegraphics[width=\columnwidth]{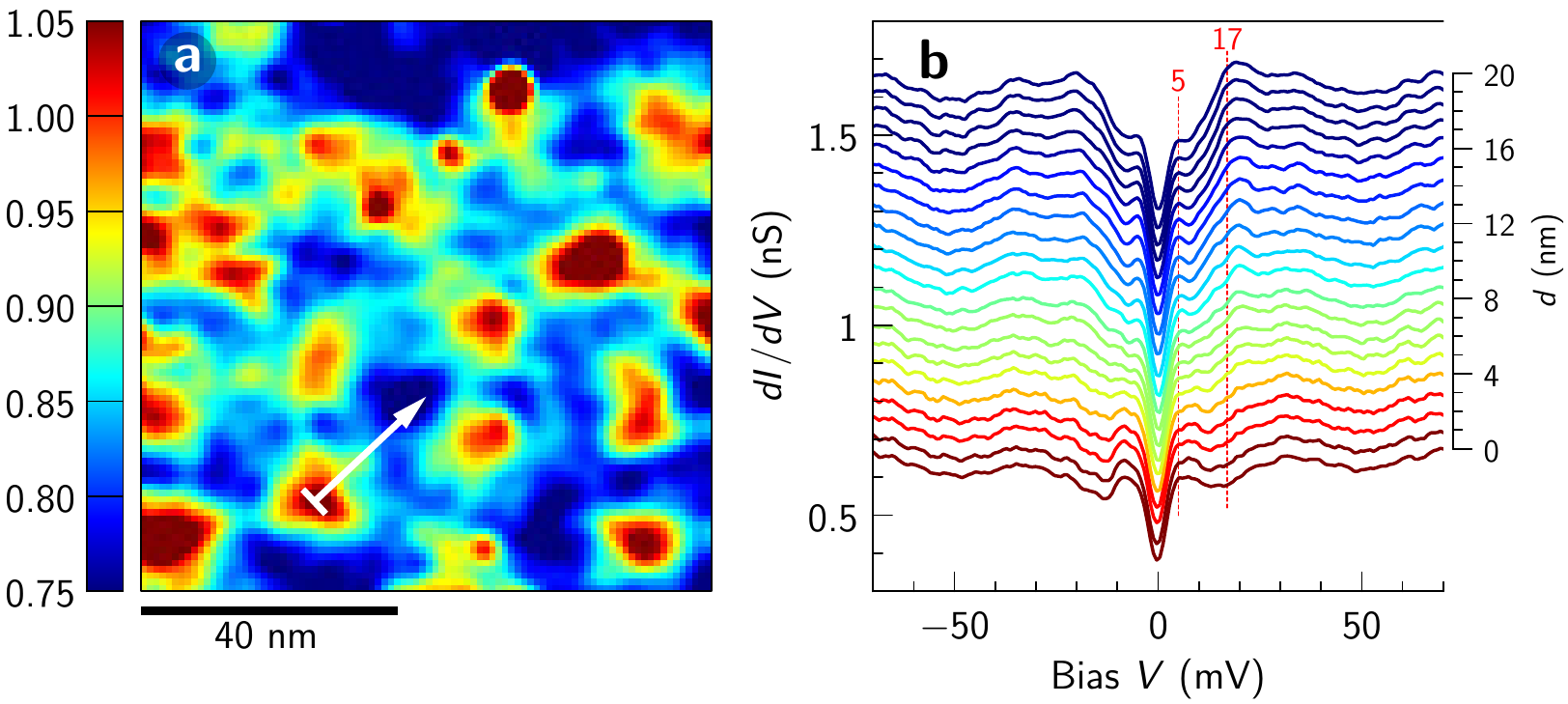}
\caption{\label{fig2} \textbf{The subgap states exist inside and outside vortices.} (\textbf{a}) Ratio $\sigma_{\mathrm{SGS}}/\sigma_{\mathrm{SC}}$ of the conductances at 5 and 17~meV measured at $T=0.4$~K in a 6~T field over a $90\times90$~nm$^2$ area. The vortex cores appear in red and the inter-vortex regions in blue. The white arrow indicates a 20~nm path starting in a vortex core. (\textbf{b}) Conductance spectra along the white arrow shown in \textbf{a}. The spectra are coloured according to the conductance ratio $\sigma_{\mathrm{SGS}}/\sigma_{\mathrm{SC}}$. The conductance scale is shown for the bottom spectrum; the others are displaced vertically by 0.05~nS for visibility. The distance $d$ to the core is shown on the right axis. The red dashed lines mark the energies 5 and 17~meV.}
\end{figure}

\begin{figure*}[tb]
\includegraphics[width=1.5\columnwidth]{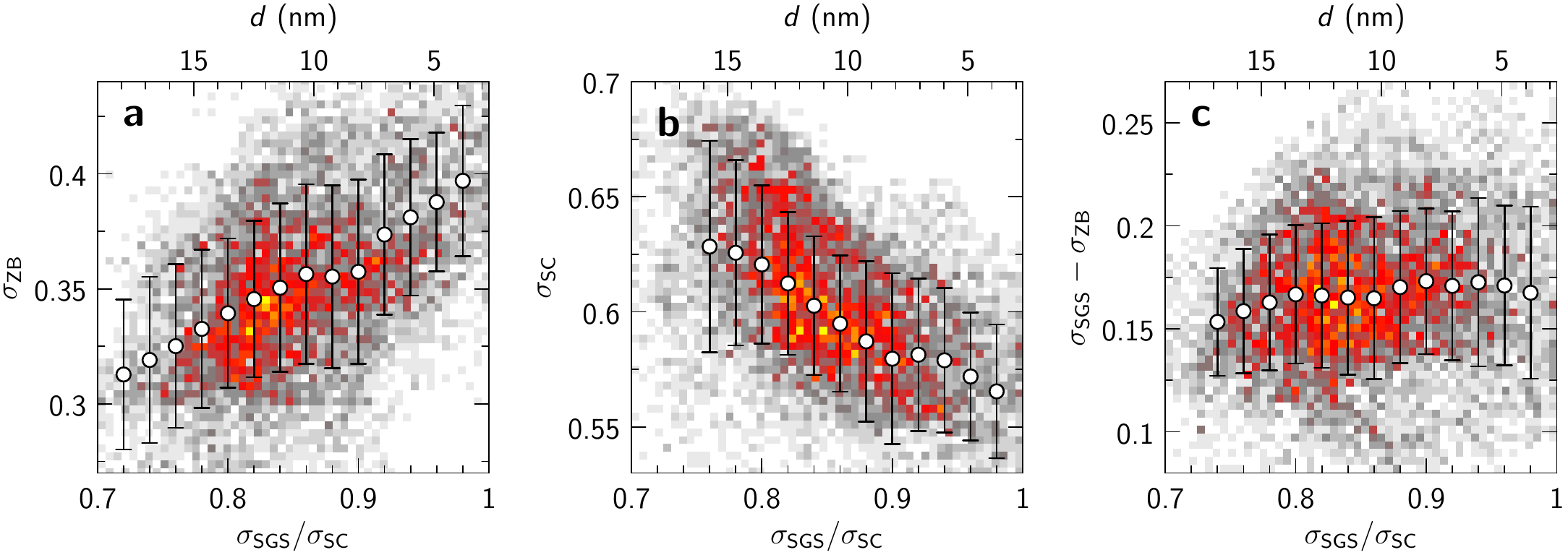}
\caption{\label{fig3} \textbf{Correlation of spectral features and distance from vortex-core centre.} The conductance ratio $\sigma_{\mathrm{SGS}}/\sigma_{\mathrm{SC}}$ evaluated along the white arrow of Fig.~\ref{fig2}a provides a measure of the distance $d$ from the vortex centre. We find a strong correlation between this distance and (\textbf{a}) the zero-bias conductance and (\textbf{b}) the conductance at $\Delta_{\mathrm{SC}}$, but not with (\textbf{c}) the amplitude of the subgap states. The colour plots are histograms showing the frequency of each pair of values on the same area as in Fig.~\ref{fig2}a (white-gray: low; red-yellow; high). The circles and error bars indicate the average and the standard deviation of the distributions at fixed values of the conductance ratio.}
\end{figure*}

Applying a field of 6~T perpendicular to the Cu-O planes, we observe vortices as shown in Fig.~\ref{fig2}a. The vortices are imaged by mapping the ratio of the conductance at 5 and 17~meV, denoted $\sigma_{\mathrm{SGS}}/\sigma_{\mathrm{SC}}$. Outside the vortices the coherence peaks are well developed and the ratio is typically 0.75; this is slightly larger than the value in zero field (0.7), because the SC peaks are smeared due to nearby vortices. Inside the vortices the ratio increases up to values above unity as the coherence peaks get totally suppressed. The spatial evolution of the conductance from the vortex core to the inter-vortex region (Fig.~\ref{fig2}b) shows unambiguously that the SGS feature is not localized inside vortices, and is the same structure as the one seen in zero field. The outstanding features of this trace are independent of direction away from the vortex core, although their relative amplitudes do depend on the proximity of neighbouring cores.

\subsection{Absence of correlation between subgap states and vortices}

Figure~\ref{fig2}b gives the visual impression that the SGS peaks are reinforced in the vortex. Upon closer inspection, this turns out to be an illusion created by the suppression of the coherence peaks in the core. In fact the pseudogap, dip, hump, and background all remain virtually unchanged upon entering the vortex. The only noticeable spectral changes along the way are an increase of the zero-bias (ZB) conductance and a suppression of the coherence peaks. The ZB conductance has a spatial distribution analogous to the conductance ratio $\sigma_{\mathrm{SGS}}/\sigma_{\mathrm{SC}}$ displayed in Fig.~\ref{fig2}a, although with a weaker contrast. This is illustrated in Fig.~\ref{fig3}a, which shows a good statistical correlation between the ZB conductance and $\sigma_{\mathrm{SGS}}/\sigma_{\mathrm{SC}}$, both increasing when approaching the vortex core at $d=0$. As expected for a weakening superconducting order parameter inside vortices, we find a clear negative correlation between the SC peak height and $\sigma_{\mathrm{SGS}}/\sigma_{\mathrm{SC}}$. In order to quantify the amount of correlation between the SGS and the vortex position, we define the amplitude of the SGS as their height above the ZB conductance, measured by the difference $\sigma_{\mathrm{SGS}}-\sigma_{\mathrm{ZB}}$. As illustrated in Fig.~\ref{fig3}c, the amplitude of the subgap states is completely uncorrelated with the vortices. This is our central result.

\section{Discussion}

\begin{figure}[b]
\includegraphics[width=\columnwidth]{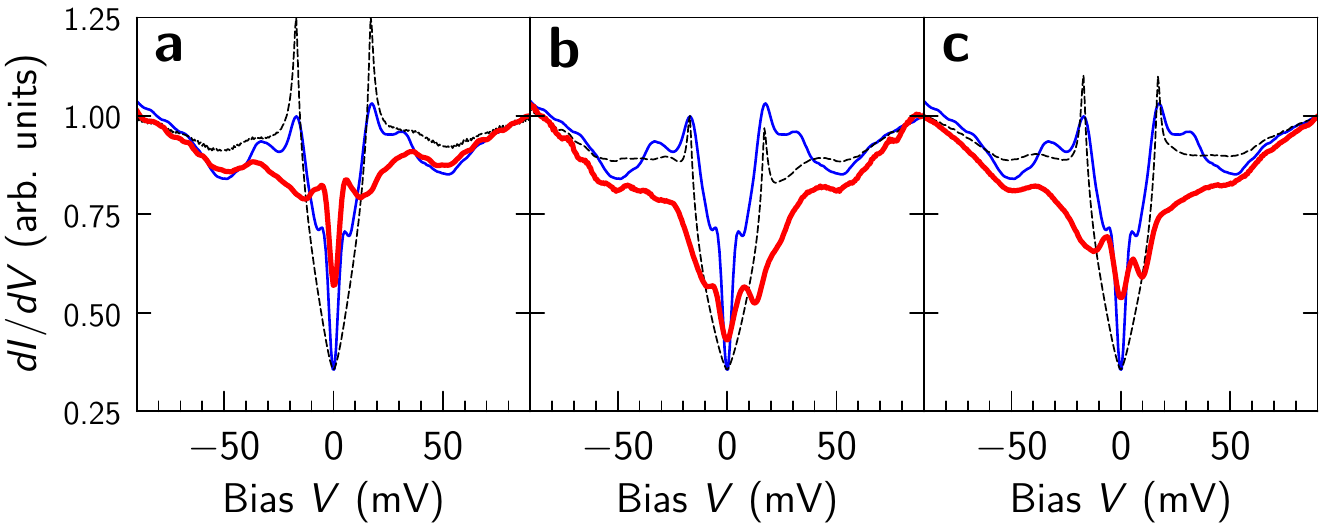}
\caption{\label{fig4} \textbf{The subgap states survive the loss of coherence.} The thick red curves show the tunnelling conductance measured at $T=0.4$~K in (\textbf{a}) a vortex core (6~T field), (\textbf{b}) at a twin boundary (zero field), and (\textbf{c}) in a non-superconducting region (zero field). The thin blue lines show a typical averaged zero-field spectrum. The dashed lines show that the model (\ref{eq}) with the red curve as $N_0(\xi)$ is inconsistent with the zero-field data; a constant was added to match the zero-bias conductance at zero-field. All curves are normalized to unity at the right edge of the graph.}
\end{figure}

The good correlation between the SC peaks and $\sigma_{\mathrm{SGS}}/\sigma_{\mathrm{SC}}$ and the absence of correlation between the SGS and $\sigma_{\mathrm{SGS}}/\sigma_{\mathrm{SC}}$ imply that there is no correlation between the SGS and the SC peak, as we have verified in the experimental data. Figure~\ref{fig4} confirms the lack of sensitivity of the SGS to the superconducting coherence. A vortex-core spectrum (Fig.~\ref{fig4}a) is compared with two zero-field spectra without coherence peaks, at a twin boundary \cite{Deutscher-1987} (Fig.~\ref{fig4}b), and inside an extended region of the sample surface where no superconducting coherence was detected (Fig.~\ref{fig4}c). All three spectra present the SGS feature at the same energy on top of different backgrounds. The PG feature appears to be also present in all three cases.

STM topography reveals atomically flat terraces (roughness below 0.2~nm) separated by full (1.2~nm) or fractional (0.4--0.8~nm) unit-cell steps as well as twin boundaries. The absence of atomic resolution prevents a positive identification of the surface layer. But it does not preclude the interpretation of the tunnelling spectra in terms of intrinsic superconducting Y123 properties as the following compelling evidences show. The gap defined by the coherence peaks satisfies the characteristic features expected for the superconducting gap of bulk Y123: its amplitude matches other bulk probes and $2\Delta_{\mathrm{SC}}/k_{\mathrm{B}}T_{\mathrm{c}}$ is in close agreement with the BCS value of 4.3 for an order parameter of $d_{x^2-y^2}$ symmetry; the coherence peaks vanish upon warming through the bulk superconducting transition temperature, in the vicinity of twin boundaries, and when entering a vortex core. The very high reproducibility of the other spectral features as a function of surface preparation, time, applied magnetic field, distance from vortex cores, step edges, and twin boundaries makes it extremely unlikely that they reflect electronic surface states. The latter are usually not as robust and one would expect modifications near step edges and twin boundaries. Lastly, there is a perfect match between the superconducting gap features measured by STM and point contact Andreev--Saint-James spectroscopy, as expected \cite{Deutscher-1999}. This body of evidences unquestionably links $\Delta_{\mathrm{SC}}$ with Y123 bulk superconductivity and rules out electronic surface states to explain the other features. We can also exclude that the SGS mark a proximity-induced or mixed $d+s$ or $d+is$ order parameter. In that case the SGS would shift to lower energy and ultimately disappear when approaching the vortex centres where phase coherence is lost, in clear contradiction with our data. Finally, the persistence of the SGS excludes their interpretation as vortex-core states or features of another state that would compete with superconductivity and pop up in the vortices.

An obvious question is whether the non-superconducting spectra shown in Fig.~\ref{fig4} represent the normal-state DOS from which superconductivity emerges. Considering an isotropic normal state (i.e., with dispersion $E_{\vec{k}}$ depending only on the modulus of $\vec{k}$) characterized by a DOS $N_0(E)$, we find that, in the presence of a $d$-wave order parameter $\Delta\cos(2\vartheta)$, the BCS DOS can be written as
	\begin{multline}\label{eq}
		N(E)=\int_{-\infty}^{\infty}d\xi\,N_0(\xi)\left(-\frac{1}{\pi}\right)\\
		\times\mathrm{Im}\left[\frac{E+i\Gamma+\xi}{\sqrt{(E+i\Gamma)^2-\xi^2}
		\sqrt{(E+i\Gamma)^2-\xi^2-\Delta^2}}\right].
	\end{multline}
The derivation is given in Supplementary Note 1. Taking the spectra of Fig.~\ref{fig4} for $N_0(\xi)$, $\Delta=17$~meV, and $\Gamma=0.5$~meV, we obtain the superconducting DOS shown as dashed lines in each panel. It is clear that the SGS are completely washed out, unlike in the experiment (Fig.~\ref{fig1}). A feature of the normal-state DOS at energy $\xi$ indeed gets spread over an energy range $[\xi,\sqrt{\xi^2+\Delta^2}]$ by the $d$-wave gap: high-energy features like the PG at $\xi\gg\Delta$ remain sharp, but low-energy features are typically smeared over a range $\Delta$. Illustrations can be found in Supplementary Figure~1.

The picture emerging from our data is that the SGS exist independently of magnetic field and are not affected by the opening of the superconducting gap. Two possible scenarios are compatible with the latter observation: the SGS either belong to an atomic layer different from the superconducting copper oxide planes or they are confined to the nodal regions in momentum space in the superconducting layers. In the first scenario the tunnelling spectrum would be a superposition of two distinct contributions: one from non-superconducting electrons characterized by the robust and homogeneous SGS as well as the ubiquitous high ZB conductance, another from the superconducting condensate responsible for the fragile coherence peaks. Indeed the spectrum of Fig.~\ref{fig1} can be perfectly reproduced by summing a pure $d$-wave spectrum and a background conductance similar to the spectra of Fig.~\ref{fig4} (see Methods). In this model the background forms the dominant spectral contribution to account for the large zero-bias and V-shaped background conductances systematically observed in Y123 tunnelling spectra. It may also explain why the vortex cores do not show the expected BCS spectrum (Figure~\ref{fig2}b). Note that our data does show an increase of the ZB conductance in the cores (Figs.~\ref{fig3}a and \ref{fig4}a), which may well be related to the predicted $d$-wave zero-bias anomaly that is obscured by the SGS. 

The second scenario assumes the SGS coexist with the superfluid condensate in the copper oxide planes. Then the SGS have to be confined to the nodal regions in momentum space where the superconducting gap is small. Otherwise they would have to shift or broaden like in Fig.~\ref{fig4} as the gap opens in the superconducting regions, in contradiction with the observations in Figs.~\ref{fig1} and \ref{fig2}. Supplementary Figure~2 presents a toy model exhibiting this property. A recent study reports particular spectroscopic signatures from the nodal regions in momentum space, however without an obvious link to the subgap states \cite{Sebastian-2014}.

In either scenario it is tempting to link the SGS with the static charge density wave (CDW) discovered recently in Y123 \cite{Ghiringhelli-2012, Chang-2012}. On one hand, the temperature scale of this phenomenon ($\sim 100$~K at optimal doping, see e.g.\ Ref.~\onlinecite{Blanco-Canosa-2014}) agrees well with the energy separation between the SGS peaks. On the other hand seemingly similar subgap structures observed in Bi2212 vortices \cite{Hoogenboom-2000a} display a modulated real-space pattern \cite{Hoffman-2002, Levy-2005, Yoshizawa-2013}. An explicit link between such conductance patterns seen by STM and the CDW was demonstrated in Bi2201 \cite{Comin-2014}. Preliminary data suggest similar real-space modulation of the SGS in Y123, but further experiments are under way.

In summary, we report the observation by STM of low-energy ($\pm$5~meV) particle-hole symmetric states in Y123. These states, seen previously only in vortex cores, are in fact a robust property of Y123, insensitive to the magnetic field and to the superconducting coherence. The subgap states often take the form of peaks in the tunnelling spectrum when the superconducting coherence peaks are absent, but thanks to high-resolution measurements we also see them as peaks even in spectra with superconducting coherence peaks. This new phenomenology redefines the question of the spectroscopy of vortices in the cuprates. Whether the new energy scale is a pairing scale, as its particle-hole symmetry suggests, whether it is connected to the Cu-O chains or has to do with some charge ordering, as suggested by similar features in Bi2212, remain open questions at this stage.

\section{Methods}

\noindent\textit{Sample preparation}.
The experiments were conducted on as-grown (001) surfaces of highly pure YBa$_2$Cu$_3$O$_{7-\delta}$ single crystals, grown in BaZrO$_3$ crucibles \cite{Erb-1995}. The samples were annealed in pure oxygen to reach an optimal doping with $T_{\mathrm{c}}=92$~K. Both as-grown and chemically etched surfaces were investigated. As-grown surfaces were simply rinsed in ethanol in an ultrasonic bath before insertion into the STM. Chemical etching was done by immersion for 1~minute in a solution of 1\% bromine and 99\% ethanol, followed by rinsing in ethanol and transfer to the vacuum chamber with less than 5 minutes exposure to air. All data reported here were measured on as-grown surfaces, except the spectra in Figs.~\ref{fig4}b and \ref{fig4}c which were measured on chemically etched surfaces. The spectra are independent of surface preparation. They are reproducible over time and after repeated thermal cycling and exposure to air on untreated surfaces, and likewise on etched surfaces if the etching is repeated.

\vspace{1em}\noindent\textit{High resolution STS measurements}.
We used a home-built high vacuum variable temperature STM optimized for high energy resolution, with a base temperature of 0.4~K, and chemically etched tips made of pure iridium. Tunnelling spectra were acquired in zero magnetic field and in a field of 6~T oriented along the surface normal. The $dI/dV$ curves, providing a measure of the electronic local density of states, were obtained by numerical differentiation of the measured $I(V)$ characteristics. For measurements done on as-grown surfaces, the junction resistance was increased to 12~G$\Omega$ when moving the tip over these surfaces (as opposed to 1.2~G$\Omega$ in the spectroscopy mode), in order to avoid that contaminants stick to the tip during its motion. For the etched surfaces the junction resistance was kept at 1.2~G$\Omega$ during scanning and spectroscopy. Both procedures result in atomically flat surfaces allowing reproducible imaging and spectroscopy.

\vspace{1em}\noindent\textit{Two-channel model for the Y123 spectrum}.
We assume that the Y123 tunnelling spectrum gets parallel contributions from a two-dimensional band distorted by the coupling to spin excitations and subject to conventional BCS $d$-wave pairing, and from a non-superconducting incoherent bath:
	\begin{equation}\label{eq:sc}
		\frac{dI}{dV}=M_dN_d(eV)+M_bN_b(eV).
	\end{equation}
$N_d(E)$ is the DOS of the BCS band and $M_d$ represents the tunneling matrix element for this band, while $M_b$ and $N_b(E)$ are the matrix element and the DOS for the non-superconducting bath, respectively. In the non-superconducting regions of the surface the coherent band turns normal with a DOS $N_0(E)$ and the differential conductance becomes
	\begin{equation}\label{eq:ns}
		\left(\frac{dI}{dV}\right)_{\mathrm{ns}}=M_d'N_0(eV)+M_b'N_b(eV)\approx
		\left(\frac{dI}{dV}\right)_{\mathrm{core}}.
	\end{equation}
The matrix elements are in general different in the two types of regions: the constant-current STM setpoint regulation may change the relative contributions of the two channels when the coherent band becomes normal and its mobility drops. We will assume that the spectrum measured in the vortex cores is a good approximation for the spectrum of non-superconducting regions. Eliminating $N_b(eV)$ from Eq.~(\ref{eq:sc}) by means of Eq.~(\ref{eq:ns}), we are left with the model
	\begin{equation}\label{eq:two-channel}
		\frac{dI}{dV}=\frac{M_b}{M_b'}\left[\left(\frac{dI}{dV}\right)_{\mathrm{core}}-M_d'N_0(eV)\right]
		+M_dN_d(eV).
	\end{equation}
\begin{figure}[t!]
\includegraphics[width=0.9\columnwidth]{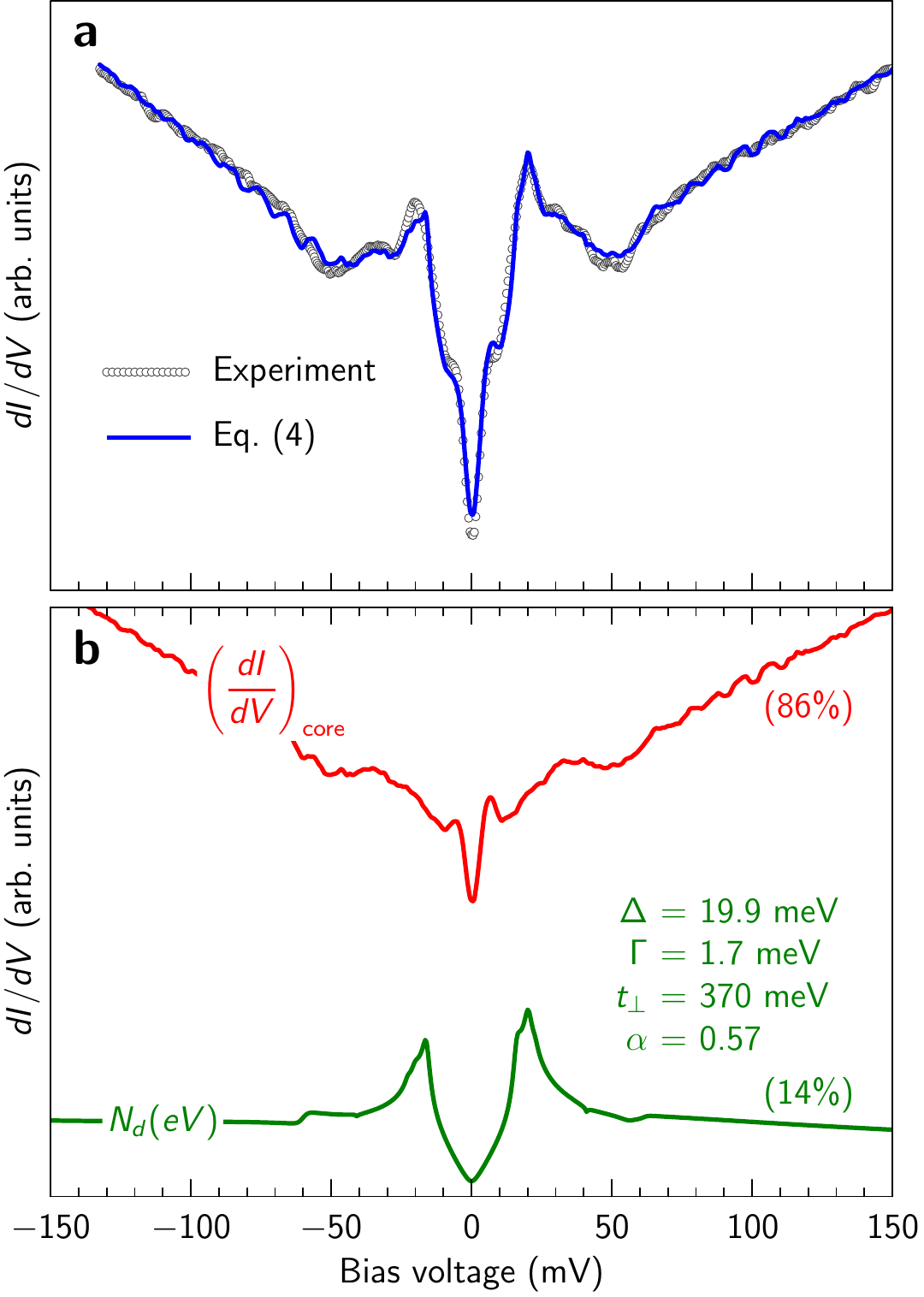}
\caption{\label{fig5} \textbf{\boldmath Two-channel model for YBa$_2$Cu$_3$O$_{7-\delta}$.} (\textbf{a}) Best fit (solid blue line) of the model (\ref{eq:two-channel}) to the tunnelling spectrum of Y123 measured between vortices (dots). (\textbf{b}) Comparison of the coherent (green) and incoherent (red) contributions to the total spectrum. The numbers in parentheses give their relative contributions to the total spectral weight in the energy range of the figure.
}
\end{figure}
We calculate $N_d(E)$ and $N_0(E)$ for a two-dimensional square lattice with a dispersion that reproduces the basic fermiology of Y123. We use the tight-binding parameters of the bonding band reported in Ref.~\onlinecite{Schabel-1998}. For simplicity we ignore the fourth and fifth neighbour hopping amplitudes and adjust the chemical potential in order to reach a hole concentration of 0.16. The resulting parameters are $(t_1,t_2,t_3,\mu)=(-281,139,-44,-356)$~meV. Y123 is a bi-layer material with both bonding and anti-bonding bands, but there is considerable uncertainty regarding the precise properties of the anti-bonding band. We add an anti-bonding band via a bilayer coupling $t_{\perp}$ and determine $t_{\perp}$ by least-square fitting. The two-band model for the dispersion is
	\begin{multline}\label{eq:BCS}
		E^{\pm}_{\vec{k}}=2t_1(\cos k_x+\cos k_y)+4t_2\cos k_x\cos k_y\\
		+2t_3(\cos 2k_x+\cos 2k_y) \pm \frac{t_{\perp}}{4}(\cos k_x-\cos k_y)^2-\mu.
	\end{multline}
The interaction with the spin resonance is included using the approach described in Ref.~\onlinecite{Eschrig-2006}. Beside its energy $\Omega_s=41$~meV, the parameters which characterize the spin resonance are its energy width $\Gamma_s$ and momentum width $\Delta q$. As the resonance is sharp in Y123 we set $\Gamma_s=0$. The resonance has a full width at half maximum 0.26~\AA$^{-1}\approx1/a$ with $a$ the in-plane lattice parameter of Y123. We therefore take $\Delta q=1/a$. The only remaining free parameter is a dimensionless coupling $\alpha$ between the Bogoliubov-de Gennes quasiparticles and the spin mode \cite{Berthod-2013}. We determine the matrix elements $M_b/M_b'$, $M_d'$, $M_d$, the model parameters $t_{\perp}$ and $\alpha$, as well as the gap $\Delta$ and Dynes broadening $\Gamma$ by least-square fitting of the expression (\ref{eq:two-channel}) to the DOS measured in-between vortices, which is very similar to the zero-field spectrum shown in Fig.~\ref{fig1}.

The result of the fit is displayed in Fig.~\ref{fig5}, where the fitted parameters are also reported. The fit yields $M_d'=0$, indicating that the coherent band is indeed invisible when it becomes normal; this supports an interpretation that this band would lie in the CuO plane below the surface and show up in the STM conductance only when it superconducts. The model captures well the asymmetric heights of the coherence peaks, the relative value of the zero-bias and background conductances, and the dips at positive and negative energies. Further refinements of the model, for instance by allowing the adjustment of all band parameters, could presumably improve the agreement even further.

\acknowledgments
We acknowledge discussions with G.\ Deutscher and S.-H.\ Pan. We thank A.\ Guipet and L.\ Stark for their technical assistance. This research was supported by the NCCR MaNEP and the Swiss National Science Foundation.

\bibliographystyle{apsrev4-1-with-titles}
\bibliography{spot,plus,supp}

\onecolumngrid
\setcounter{equation}{0}
\renewcommand{\theequation}{S\arabic{equation}}
\setcounter{figure}{0}
\renewcommand{\thefigure}{S\arabic{figure}}

\section*{Supplementary Note 1}

\twocolumngrid

\noindent\textit{One-channel models for the spectrum of Y123 in zero field}.
One-channel models assume that the tunnelling conductance measured by STM on the surface of Y123 gets contributions from a single electronic band of dispersion $\xi_{\vec{k}}=\varepsilon_{\vec{k}}-\mu$, on top of a featureless background associated with transport channels of the as-grown surface and responsible for the large zero-bias conductance. In this frame of mind, the subgap peaks represent a property of the normal state which participates in pairing and is therefore gapped in the superconducting state. The dispersion $\xi_{\vec{k}}$ defines a normal-state density of states (DOS)
	\begin{equation}\label{eq:N0}
		N_0(E)=\int\frac{d^2k}{(2\pi)^2}\delta(E-\xi_{\vec{k}}).
	\end{equation}
The opening of a superconducting gap changes the DOS into
	\begin{equation}\label{eq:N}
		N(E)=-\frac{1}{\pi}\int\frac{d^2k}{(2\pi)^2}\mathrm{Im}\left[\frac{E+i\Gamma+\xi_{\vec{k}}}
		{(E+i\Gamma)^2-\xi_{\vec{k}}^2-\Delta_{\vec{k}}^2}\right].
	\end{equation}
The quantity in brackets is the retarded single-particle Green's function of a BCS superconductor with a gap function $\Delta_{\vec{k}}$, and $\Gamma$ is a Dynes broadening parameter representing residual impurity scattering. We restrict ourselves to two-dimensional models.

\vspace{1em}\noindent\textit{Arbitrary isotropic normal state}.
We first consider a normal-state dispersion that is isotropic in the plane, $\xi_{\vec{k}}=\xi_{|\vec{k}|}\equiv\xi(k)$, but otherwise arbitrary. The Fermi surface is circular, and the details of the normal-state DOS $N_0(E)$ are controlled by the dependence of the dispersion on the modulus of $\vec{k}$. Any function $N_0(E)$ can in principle be parametrized by a monotonically increasing radial dispersion $\xi(k)$, by solving the implicit equation $2\pi N_0(\xi)\xi'(k)=k(\xi)$. Our goal is to express the superconducting DOS (\ref{eq:N}) in terms of the arbitrary function $N_0(E)$. For an $s$-wave gap, this is straightforward:
	\begin{equation}\label{eq:Ns}
		N_s(E)=\int_{-\infty}^{\infty}d\xi\,N_0(\xi)\,\mathrm{Im}
		\left[\frac{-\frac{1}{\pi}(E+i\Gamma+\xi)}{(E+i\Gamma)^2-\xi^2-\Delta^2}\right].
	\end{equation}
Hence for any normal-state DOS function $N_0(E)$, the one-channel $s$-wave superconducting DOS for a gap magnitude $\Delta$ can be computed by performing numerically the integral in (\ref{eq:Ns}). If the superconducting state has $d$-wave symmetry, the situation is more complicated because of the angular dependence of the gap. An explicit expression can nevertheless be obtained if the gap has the functional form $\Delta\cos2\vartheta$. Let us write $N_d(E)=-(1/\pi)\,\mathrm{Im}\,N_d(z\to E+i\Gamma)$ with
	\[
		N_d(z)=\int\frac{d^2k}{(2\pi)^2}\frac{z+\xi(k)}{z^2-\xi^2(k)-(\Delta\cos2\vartheta)^2}.
	\]
Working in cylindrical coordinates, we note that
	\begin{align*}
		&\int_0^{2\pi}d\vartheta\,\frac{z+\xi(k)}{z^2-\xi^2(k)-(\Delta\cos2\vartheta)^2}\\
		&\qquad=2\pi\frac{z+\xi(k)}{\sqrt{z^2-\xi^2(k)}\sqrt{z^2-\xi^2(k)-\Delta^2}}\\
		&\qquad=\int_0^{2\pi}d\vartheta\,\frac{z+\xi(k)}{\sqrt{z^2-\xi^2(k)}\sqrt{z^2-\xi^2(k)-\Delta^2}}.
	\end{align*}
Thus $N_d(z)$ can be rewritten as a two-dimensional integral without angular momentum in the integrand, and then expressed in terms of the normal-state DOS:
	\begin{align}\label{eq:Nd}
		\nonumber
		N_d(z)&=\int\frac{d^2k}{(2\pi)^2}\frac{z+\xi(k)}{\sqrt{z^2-\xi^2(k)}\sqrt{z^2-\xi^2(k)-\Delta^2}}\\
		&=\int_{-\infty}^{\infty}d\xi\,N_0(\xi)
		\frac{z+\xi}{\sqrt{z^2-\xi^2}\sqrt{z^2-\xi^2-\Delta^2}}.
	\end{align}
This leads to Eq.~(1) of the main text.

\begin{figure}[b]
\includegraphics[width=0.75\columnwidth]{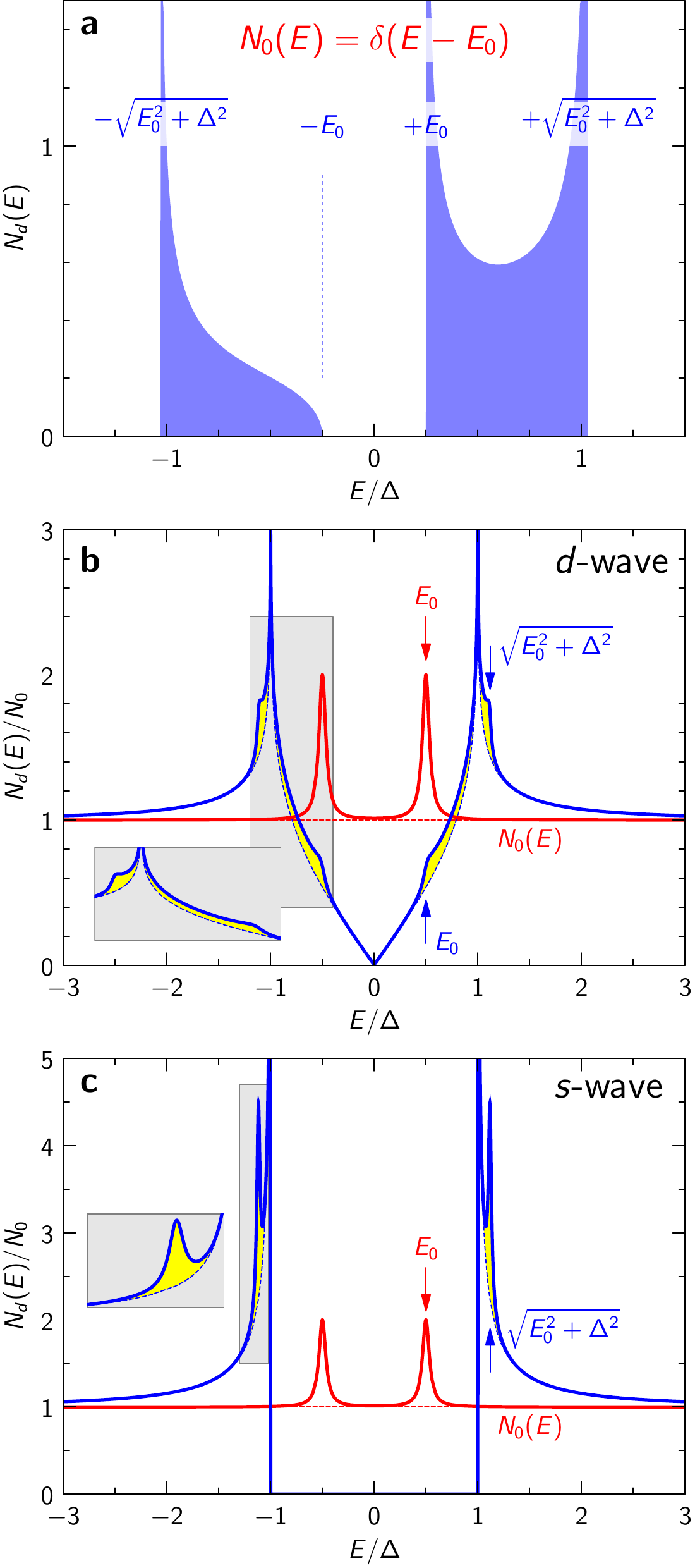}
\caption{\label{fig:singularities}
\textbf{\boldmath Spreading of a normal-state subgap peak by the opening of a $d$-wave gap.}
(\textbf{a}) DOS calculated with Supplementary Equation~(\ref{eq:Nd}) for a normal-state DOS $N_0(E)=\delta(E-E_0)$ with $E_0/\Delta=0.25$. (\textbf{b}) $d$-wave and (\textbf{c}) $s$-wave superconducting DOS (blue curves) for a normal-state DOS $N_0(E)$ given by a constant background with two symmetric peaks at $E_0=\pm\Delta/2$ (red curves). The dashed lines correspond to a constant normal-state DOS $N_0(E)=N_0$. The yellow area highlight the spectral weight of the peaks in the superconducting DOS, and the insets show a zoom of the peak region.
}
\end{figure}

Supplementary Equation (\ref{eq:Nd}) shows that if the normal-state DOS has sharp structures, these structures are spread by the opening of the $d$-wave gap. To see this, take $N_0(\xi)=\delta(\xi-E_0)$ and $\Gamma=0^+$ in (\ref{eq:Nd}). The resulting superconducting DOS has three square-root divergences at $E=E_0$ and $E=\pm(E_0^2+\Delta^2)^{1/2}$ and a square-root singularity at $E=-E_0$. The spectral weight of the delta peak is spread into the intervals between these two couples of singularities [Supplementary Figure~\ref{fig:singularities}(a)]. The case of a normal-state DOS that is the sum of a flat background and two symmetric peaks at $E=\pm E_0$ is illustrated in Supplementary Figure~\ref{fig:singularities}(b) for $d$-wave pairing and Supplementary Figure~\ref{fig:singularities}(c) for $s$-wave pairing. While the peak structure remains visible in the $s$-wave case, nothing but weak features subsist in the $d$-wave case. In the main text it is shown that this one-channel model with an isotropic normal-state dispersion cannot explain the subgap peaks present in the STM spectra of Y123. A weak anisotropy of the dispersion will not change this conclusion qualitatively.

\vspace{1em}\noindent\textit{Strongly anisotropic normal state with nodal peaks}.
The failure of the one-channel isotropic model to account for the tunnelling spectrum of Y123 suggests to devise a strongly anisotropic model in which the two low-energy peaks in the normal-state DOS would be related to features localized near the nodal directions in reciprocal space. These features would not be broadened by the opening of the $d$-wave gap and could survive in the superconducting spectrum. Such a model is somewhat contradictory. On the one hand, we need a structure with a well-defined energy to produce prominent peaks in the normal-state. This is usually associated with a localized state, a non-dispersing band, or any phenomenon with no dispersion in momentum space. But the absence of dispersion in momentum space is the quintessence of isotropy!

We can escape this paradox if the two symmetric peaks in the normal-state DOS are due to a gap in the normal-state dispersion, which vanishes away from the nodal regions. We therefore introduce a normal-state gap $\Theta_{\vec{k}}=E_0\theta(w-|\cos k_x-\cos k_y|/2)$, where $\theta$ is the Heaviside function. The parameter $w$ controls the extension of the gapped region around the nodal lines, $w=0$ corresponding to no gap and $w=1$ to a fully gapped Fermi surface. The modified normal-state dispersion is $\bar{\xi}_{\vec{k}}=\mathrm{sign}(\xi_{\vec{k}})(\xi_{\vec{k}}^2+\Theta_{\vec{k}}^2)^{1/2}$. We try this model on a two-dimensional square lattice with the same dispersion as for the two-channel model presented in the Methods section, Eq.~(5), but without interlayer coupling for simplicity ($t_{\perp}=0$). The corresponding Fermi surface is shown in Supplementary Figure~\ref{fig:nodalpeaks}(b). Using the gapped dispersion $\bar{\xi}_{\vec{k}}$ with $E_0=5$~meV and a superconducting gap $\Delta_{\vec{k}}=\Delta(\cos k_x-\cos k_y)/2$ with $\Delta=19$~meV, we calculate the normal and superconducting DOS using Supplementary Equations~(\ref{eq:N0}) and (\ref{eq:N}) and $\Gamma=0.1$~meV. The results are displayed in Supplementary Figures~\ref{fig:nodalpeaks}(a) and \ref{fig:nodalpeaks}(c).

The model with $w=0.1$ indeed has two peaks from the normal-state DOS surviving in the superconducting DOS. A new energy scale appears below $E_0$. This is because the superconducting gap at the edges of the gapped region is smaller than $E_0$. Note that our choice of the function $\Theta_{\vec{k}}$ with the same functional form as $\Delta_{\vec{k}}$ implies that the edges of the gapped region are iso-contours of $\Delta_{\vec{k}}$. At these edges we have simply $\Delta_{\vec{k}}=w\Delta=1.9$~meV. There are no Bogoliubov quasiparticles with energy lower than $w\Delta$, such that the superconducting DOS is completely gapped below this energy. Between $w\Delta$ and $E_0$, the momenta outside the gapped region contribute to the DOS, which approaches the $w=0$ curve from below, due to the missing states from the gapped regions. At $E_0$ we have the remnant of a normal-state square-root singularity of weight $\sim w$ due to the nearly flat dispersion of states inside the gapped region, and above $E_0$ the superconducting dispersion returns to conventional, and so does the DOS.

For $w=0.5$, the DOS vanishes below $E_0$ because $w\Delta>E_0$ and $E_0$ is the smallest gap on the Fermi surface. There is no peak at $E_0$, the energy of the gapped states near the nodes being spread between $E_0$ and $[E_0^2+(w\Delta)^2]^{1/2}$, where a weak peak remains. As seen in Supplementary Figure~\ref{fig:nodalpeaks}(c), the weight of the in-gap peak is independent of $w$, despite the fact that in the normal-state DOS this weight is proportional to $w$.

The model of Supplementary Figure~\ref{fig:nodalpeaks} is as a proof of principle, but we consider unlikely that it explains the subgap states in Y123. The latter have a spectral weight comparable with the superconducting coherence peaks, while in Supplementary Figure~\ref{fig:nodalpeaks} the weight of the in-gap peaks is tiny. More importantly, the measurements show that this weight remains the same in the superconducting and non-superconducting states, which is not the case in the model.

\begin{figure}[b]
\includegraphics[width=\columnwidth]{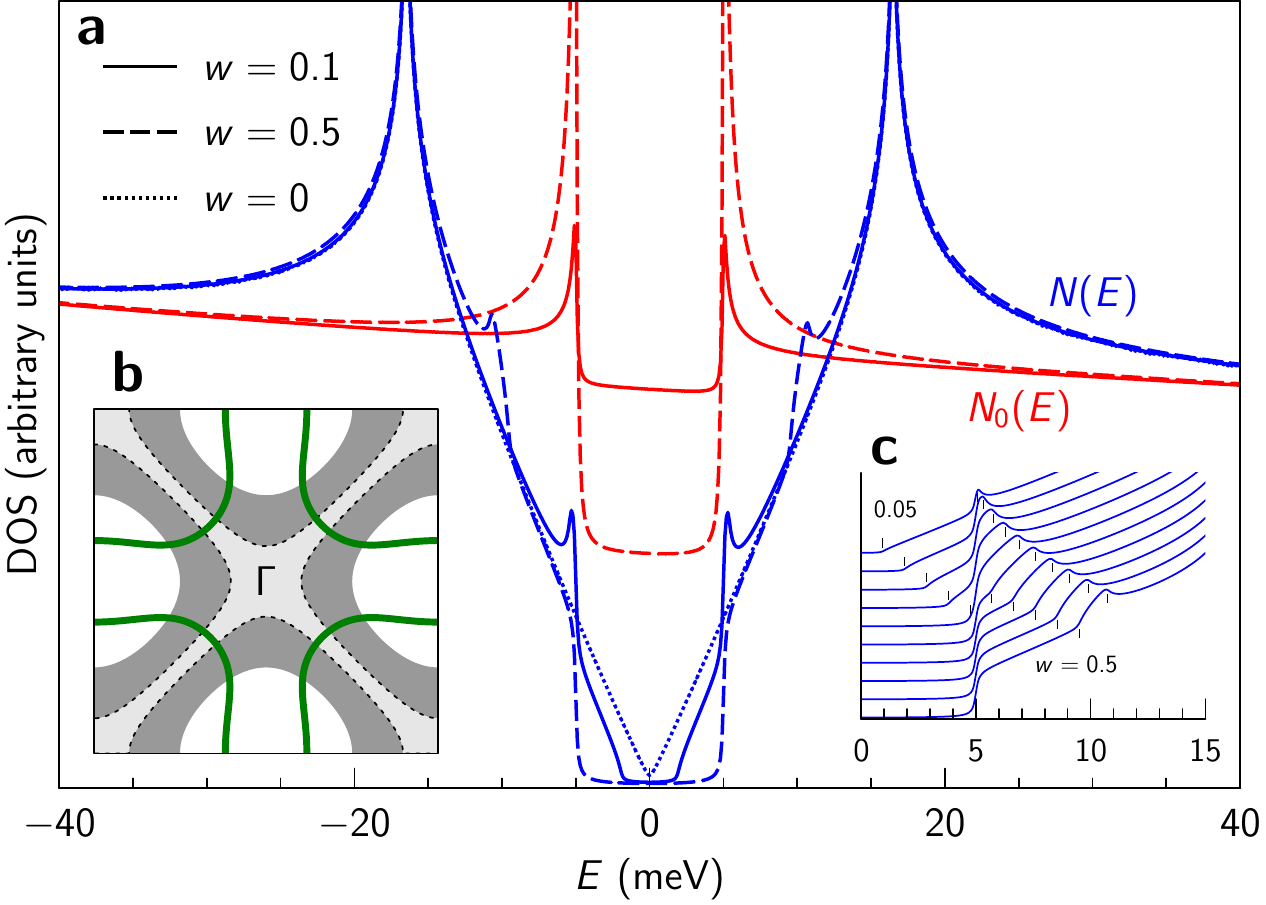}
\caption{\label{fig:nodalpeaks}
\textbf{\boldmath Toy model with gapped Fermi surface in the nodal regions.}
(\textbf{a}) Normal (red) and superconducting (blue) DOS for a model with gapped normal-state dispersion along the nodal directions. The solid lines correspond to $w=0.1$, the dashed ones to $w=0.5$, and the dotted line shows the usual $d$-wave DOS for $w=0$. (\textbf{b}) Fermi surface (green) and gapped region around the nodal lines for $w=0.1$ (lightgray) and $w=0.5$ (gray). The thickness of the green line covers all momenta contributing to the DOS in \textbf{a}. (\textbf{c}) Low-energy superconducting DOS for values of $w$ varying between $0.05$ and $0.5$; the curves are offset vertically. The bars indicate the energies $w\Delta$ and $[E_0^2+(w\Delta)^2]^{1/2}$.
}
\end{figure}

\end{document}